\documentclass[aps,pra,twocolumn,groupedaddress,showpacs]{revtex4}
\usepackage{multirow}
\usepackage{graphicx}

\begin{document}
\title{High-performance diamond-based single-photon sources for quantum communication}

\author{Chun-Hsu Su}
\email{chsu@ph.unimelb.edu.au}
\affiliation{School of Physics, University of Melbourne, VIC 3010, Australia}

\author{Andrew D. Greentree}
\affiliation{School of Physics, University of Melbourne, VIC 3010, Australia}
\author{Lloyd C. L. Hollenberg}
\affiliation{Centre for Quantum Computer Technology, School of Physics, University of Melbourne, VIC 3010, Australia}

\date{\today}
\begin{abstract}
Quantum communication places stringent requirements on single-photon sources. Here we report a theoretical study of the cavity Purcell enhancement of two diamond point defects, the nickel-nitrogen (NE8) and silicon-vacancy (SiV) centers, for high-performance, near on-demand single-photon generation. By coupling the centers strongly to high-finesse optical photonic-bandgap cavities with modest quality factor $Q = O(10^4)$ and small mode volume $V = O(\lambda^3)$, these system can deliver picosecond single-photon pulses at their zero-phonon lines with probabilities of 0.954 (NE8) and 0.812 (SiV) under a realistic optical excitation scheme. The undesirable blinking effect due to transitions via metastable states can also be suppressed with $O(10^{-4})$ blinking probability. We analyze the application of these enhanced centers, including the previously-studied cavity-enhanced nitrogen-vacancy (NV) center, to long-distance BB84 quantum key distribution (QKD) in fiber-based, open-air terrestrial and satellite-ground setups. In this comparative study, we show that they can deliver performance comparable with decoy state implementation with weak coherent sources, and are most suitable for open-air communication.
\end{abstract}


\maketitle 

\section{Introduction}
Light sources that reliably prepare single photons in well-defined spatiotemporal modes on demand are enablers for experiments in quantum optics~\cite{walls95}, precision optical measurement~\cite{polzik92,giov04} and numerous prospective quantum information technologies such as quantum teleportation~\cite{kwiat95}, quantum repeaters~\cite{duan01}, and linear-optical quantum computers~\cite{knill01}. Their most immediate application is in quantum cryptography~\cite{gisin02}. The faithful and practical implementation of the well-known BB84 quantum key distribution (QKD) scheme~\cite{bennett84} relies on the use of a high-performance true single-photon source (SPS). Most QKD systems realized to date employ weak coherent sources (WCS) with relatively low single-photon throughput. While the resultant key creation rate can be improved drastically with decoy states~\cite{hwang03,wang05,lo05}, further enhancement and simplification of practical BB84 implementations are possible using SPSs~\cite{horikiri06}. 

There are numerous schemes for single-photon generation including two-photon down-conversion~\cite{hong85}, Coulomb blockade of electrons~\cite{kim98}, state reduction~\cite{varcoe99}, and fluorescence of a wide range of quantum systems, e.g. atoms~\cite{hijlkema07}, trapped ions~\cite{keller04}, molecules~\cite{brunel99,lounis00}, and quantum dots~\cite{shields07}.
Various optically-active point defects in diamonds are also superb SPS candidates and have been extensively investigated. In particular, antibunching experiments with isolated centers, namely the negatively-charged nitrogen-vacancy (NV)~\cite{kurtsiefer00}, nickel-nitrogen (NE8)~\cite{gaebel04, wu07} and silicon-vacancy (SiV)~\cite{wang06} centers, have also found them to be photostable at room temperature and robust against photobleaching. Clock-triggered NV and NE8 based SPSs were demonstrated in Refs.~\cite{beveratos02,alleaume04} (for QKD) and \cite{wu07} respectively. Coupling of a diamond nanocrystal containing NV to an optical fiber has also been reported~\cite{kuhn01,rabeau05} and fiber-coupled sources are now commercially available~\cite{trpkovski08}. 

Despite these successes, the bare centers are not ideal for practical QKD and prospective technologies. Their emission is nondirectional and this impedes collection efficiency. The nanosecond photoluminescence lifetimes of these centers limit their theoretical operational speeds to $< 500$~MHz. The presence of a metastable state leads to undesirable blinking and $>100$~ns long dark intervals, and decays via upper-phonon transitions lead to broadened emission bandwidths~\cite{gaebel04, wang06, neumann09}. Also, the SiV center suffers from fast nonradiative decay that degrades its fluorescence~\cite{feng93} and ultimately reliability. Although there are other newer, uncharacterized defects~\cite{igor09,simpson09,igor09b} with large transition dipole moments, we focus on these most commonly discussed centers.

To realize a directional and high-performance source, the emitter can be placed inside an optical resonator. In this cavity quantum electrodynamic (QED) arrangement, coupling between an NV center and an active cavity mode has been studied previously in Ref.~\cite{su08}, showing dramatic improvements in the spectral purity and emission rate. In the first part of this work (Sec. II), we study the use of cavities to enhance the emission of the NE8 and SiV centers -- but with a clear distinction owing to appreciable differences in their electronic and vibronic structures. We show that wavelength-sized cavities with quality factor $Q = O(10^4)$ improve their emission rates by over two orders of magnitude, and overcome the photoluminescence problem of the SiV center and the common blinking effect. In Sec.~III, we explore the potential of these sources (including the NV) toward various QKD implementations. Among silica fiber-based, open-air terrestrial and satellite-ground arrangements, we find that the cavity-enhanced centers deliver performance comparable with the decoy state, and are most suitable for open-air communication.

\begin{figure}[tb!]
\includegraphics[width=0.85\columnwidth]{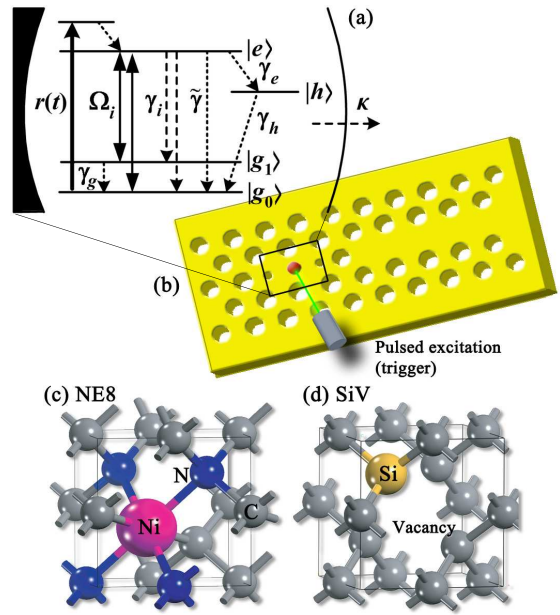}
\caption{(Color online) (a) Schematic of a cavity-center system for single-photon generation. The NE8 and SiV centers are modelled as a vibronic system that consists of a single excited state $|e\rangle$ and a ground state with two vibrational sublevels $|g_i\rangle$. $|h\rangle$ is an additional metastable level. The center is pumped with an external classical field $r(t)$ acting as the trigger pulse. The $|e\rangle-|g_i\rangle$ transition is coupled to a one-sided, single-moded cavity with coupling strength $\Omega_i$. Dashed arrows denote radiative decay and dotted arrows non-radiative relaxations.
(b) Proposed design in diamond-based PBG lattice, missing holes constitute the cavities and waveguide. Solid circle (in box) denotes a center excited optically by a green laser. (c) NE8 complex in the carbon lattice, consisting of a Ni atom surrounded by four N atoms. (d) SiV structure consisting a substitutional Si atom and an adjacent vacancy.}
\label{fig:schematics}
\end{figure}

Photonic crystals (PCs) are ideally suited for realizing the required cavity properties. The schematic in Fig.~\ref{fig:schematics}(b) envisages such an implementation. A defect in a planar photonic-bandgap (PBG) structure defines an excellent cavity, and extremely high-$Q/V$ cavities [mode volume $V = O(\lambda^3), Q = O(10^6)$] for near-infrared (near-IR) operation have been demonstrated in silicon~\cite{noda07}. For enhancing the visible emission from these centers, compatible high-$Q/V$ cavity designs are also available in diamond~\cite{tom06,bayn07,kreuzer08,bayn08,tom09} and gallium phosphide (GaP)~\cite{barclay09}. Although fabrication of cavity-enhanced centers in integrated photonics~\cite{greentree08} remains challenging, experimental demonstration of diamond-based PBG cavity of $Q = 585, V = O(\lambda^3)$~\cite{wang07} and coupling an NV to a silicon nitride cavity of $Q = 3200$~\cite{barth09}, give cause for optimism. Realizing a coupled NV-cavity system has also been achieved with $Q > 25000$ microdisk in GaP~\cite{barclay09b}. Finally, we note that novel slot-waveguide geometries, combined with mirrors, PBG or distributed Bragg reflectors in a Fabry-P\'{e}rot arrangement, are promising for achieving ultra-high optical confinement~\cite{almeida04,rob05,hiscocks09}.

\section{Cavity-enhanced diamond centers}
Diamond color centers are vibronic systems that consist of a ground state manifold of vibrational sublevels $\{|g_i\rangle\}$ and an excited state manifold of sublevels $\{|e_j\rangle\}$, where $i, j = 0, 1, 2,...$ label these levels with increasing energy. Transitions between the manifolds are electric dipole in nature, giving rise to multiple phonon lines, denoted $i$PL, in the emission spectrum: $|e_0\rangle\rightarrow|g_0\rangle$ transition is the sharp zero-phonon line (ZPL), and the vibronic sidebands due to phonon-assisted transitions $|e_0\rangle\rightarrow|g_k\rangle$ ($k>0$) are broadened by rapid phononic relaxations. Below, we rename $|e_0\rangle$ as $|e\rangle$ as shorthand. The NE8 center has only very weak sidebands at room temperature and a prominent ZPL centered at $\lambda_{\rm ZPL} = 794$~nm. The relative integrated intensity of the ZPL to the entire spectrum is 0.7~\cite{gaebel04}. For the SiV center, its ZPL centered at 738~nm makes up $\sim 80\%$ of the emission intensity~\cite{clark95}. Their 1PLs are roughly centered at 20--30~nm from the ZPLs. Thus, we adopt a common three-state model $\{|g_0\rangle, |g_1\rangle$, $|e\rangle\}$ with photoluminescence lifetimes $\tau$ of 11.5~ns (NE8) and 2.7~ns (SiV). Added to our model is a nonradiative metastable state $|h\rangle$ that is responsible for dark intervals (blinking) and the photon-bunching signature in the autocorrelation function $g^{(2)}$ of their emissions~\cite{gaebel04, wang06}. The complete model inside a one-sided, single-moded cavity is depicted in Fig.~\ref{fig:schematics}.

In dipole and rotating-wave approximations, the Hamiltonian of the joint center-cavity system is
\begin{equation}
 \mathcal{H}/\hbar = \sum_{\alpha} \omega_{\alpha}\sigma_{\alpha\alpha} + \omega_c a_c^\dagger a_c + 
 \sum_{i=0, 1}\Omega_{i}(a_c^\dagger \sigma_{g_i e} + {\rm h.c.}),
 \label{eq:ham}
\end{equation}
where $\sigma_{\alpha\beta} = |\alpha\rangle\langle\beta|$ is the atomic projection operator, $|\alpha\rangle$ and $|\beta\rangle$ denote the atomic states, and $a_c^\dagger$ ($a_c$) is the creation (annihilation) operator for the cavity mode-$c$. $\hbar\omega_\alpha$ is the energy of the atomic state $|\alpha\rangle$ and $\hbar\omega_c = 2\pi \hbar c/\lambda_c$ is the resonant energy of the cavity. The dipole transitions $i$PL are coupled to the cavity mode with single-photon Rabi frequency $\Omega_i = d_i[\omega_c/(2\hbar\epsilon_0 V)]^{1/2}$, parameterized by the transition dipole moment $d_i$ and the cavity mode volume $V$. 

The dynamics of the system coupled to an adjacent waveguide (mode-$w$) satisfies the master equation for its density operator $\rho$, 
\begin{eqnarray}
	\dot{\rho} & = & -\frac{i}{\hbar}[\mathcal{H},\rho] + \gamma_e\mathcal{L}[\sigma_{he}] + \gamma_h\mathcal{L}[\sigma_{g_0h}] + \gamma_g\mathcal{L}[\sigma_{g_0 g_1}] \\
	& + & \sum_{i=0,1}\gamma_{i}\mathcal{L}[\sigma_{g_ie}] + \tilde{\gamma}\mathcal{L}[\sigma_{g_0e}] + \kappa\mathcal{L}[a_w^\dagger a_c] + r(t)\mathcal{L}[\sigma_{eg_0}], \nonumber
\label{eq:mastereqn}
\end{eqnarray}
where the Liouvillean, defined for an operator $A$,
\begin{equation}
	\mathcal{L}[A] \equiv A\rho A^\dagger - \frac{1}{2}\Big(A^\dagger A\rho + \rho A^\dagger A\Big),
\end{equation}
is used to describe the relevant atomic relaxation processes, namely the relaxation to the metastable state of lifetime $1/\gamma_h$ with rate $\gamma_e$, the phononic relaxation within the manifold with rate $\gamma_g \gg \gamma_i$, the atomic (radiative) spontaneous transitions $i$PL with decay rate $\gamma_i$, and other nonradiative decay processes with combined rate $\tilde{\gamma}$. $\kappa \equiv \omega_c/(2Q)$ is the rate of cavity loss into mode-$w$ associated with the creation operator $a_w^\dagger$. We assume that this effect dominates the determination of $Q$. Finally, we include a pulsed excitation with pumping rate $r(t)$ that serves as a trigger for single-photon emission. The magnitude of the transition dipole moment $d_i$ of $i$PL transition can be determined using the relation $d_i^2 = 3\pi\hbar\epsilon_0 c^3/(\gamma_i n \omega_i^3)$, where $n\sim 2.4$ is the refractive index of diamond and $\hbar\omega_i$ is transition energy. 
Allowing the possibility of multiple excitations/de-excitations per pulse, we adopt a basis set $\{|e\rangle, |g_0\rangle, |g_1\rangle, |h\rangle\} \otimes \{|0_c\rangle, |1_c\rangle,..., |N_c\rangle\} \otimes \{|0_w\rangle, |1_w\rangle,..., |N_w\rangle\}$, where $|n_c\rangle$ ($|n_w\rangle$) denotes the number state of mode-$c$ ($w$).

Although the shelving level reduces photon emission, the quantum yield of the bare NE8 center remains near unity, i.e. $\tilde{\gamma} = 0$. In comparison, the SiV center has a very low quantum yield of $\sim0.05$ at room temperature~\cite{turukhin96} largely due to nonradiative transitions so that $\tilde{\gamma}=19/\tau$. Since the branching ratio $\gamma_0/\gamma_1$ is related to the relative integrated intensity of the ZPL to the total emission, we take $\gamma_0/\gamma_1 = 7/3$ (NE8) and 4 (SiV). Fitting to the measured autocorrelation function $g^{(2)}$ in Ref.~\cite{gaebel04} gives $\gamma_e/2\pi =17$~MHz and $\gamma_h/2\pi = 6.1$~MHz for the NE8. For the SiV, $\gamma_e/2\pi \sim 30$~MHz and $\gamma_h/2\pi \sim 10$~MHz~\cite{wang06}. In this model, we have explicitly ignored thermal broadening but it can be introduced phenomenologically for a more realistic estimate of the spectral width of the output wave packet.

Reliable single-photon generation from the cavity-center system requires that losses via atomic relaxation and the shelving state be minimized, thus the atomic excitation should be outcoupled into mode-$w$ faster than these processes. In the strong Purcell regime~\cite{law97} where the cavity is resonant with the $M$th phonon transition and $\kappa > \Omega_M \gg \{\gamma_i (\forall_i), \gamma_e, \gamma_h, \tilde{\gamma}\}$, the Rabi frequency dominates the rates of the incoherent processes so that the state vector evolves as $|e,0_c,0_w\rangle \rightarrow |g_M,1_c,0_w\rangle \rightarrow |g_M,0_c,1_w\rangle$. Notably, $\kappa > \Omega_M$ ensures rapid outcoupling from the cavity mode and suppresses the vacuum Rabi oscillations, which otherwise lead to unwanted Rabi remnants -- ``ringing'' spectral features at the cavity output. 

Here we focus on enhancing the ZPL transition with diamond-based PBG cavities of $V=(\lambda_c/n)^3$, which is of the order typically obtained in proposed cavity designs~\cite{tom09}. Although the theoretical limit for the fundamental mode of such a cavity is $(\lambda_c/2n)^3$, these modes are normally not used because they are lossy~\cite{noda07}. For the upper phonon transitions to be enhanced, a large coupling $\Omega_M \gg \gamma_g$ and hence sub-wavelength confinement [$V \ll (\lambda_c/n)^3$] are required due to their associated broad linewidth~\cite{su08}.

Before presenting the simulation results, the optimal operating regime warrants further discussion. The cavity enhancement of a quantum emitter is measured by the Purcell factor $F_p$, defined as the ratio of the emission rate to the cavity mode to the no-cavity case~\cite{purcell46,haroche85,cui05}. In this case, it is given by
\begin{equation}
	F_p = \frac{4\Omega_0^2}{\gamma_{\rm total}\kappa} = \frac{4d_0^2}{\hbar\epsilon_0 \gamma_{\rm total}}\frac{Q}{V}
	\label{eq:purcell}
\end{equation}
where $\gamma_{\rm total} = 2\pi/\tau+\tilde{\gamma}+ \gamma_e$ is the total decay rate from state $|e\rangle$ expressed in units of angular frequency. Purcell enhancement is maximized with a high-$Q/V$ cavity (i.e. small $\kappa$), but an arbitrarily large $Q$ delays cavity output. For a passive structure, there is an upper bound for $Q$, or equivalently, $\kappa \geq 2\Omega_0$. Hence the $Q$ factor is optimal when $\kappa \approx 2.5\Omega_0$ and the Rabi remnants suppressed.

When the center is pumped too strongly, the use of cavities makes multi-photon emission events per trigger possible. The excitation parameter $r(t)$ must be optimized to maximize excitation probability and single-photon emission probability $P_1$ while suppressing multi-photon probability $P_{\rm m}$. In general, the temporal extent of an excitation pulse must be much shorter than the time scales for Rabi oscillation and cavity decay. During excitation we can then ignore the effect of decoherence and treat the centers are two-level systems. Assuming a top-hat trigger pulse of length $T$ and absorption rate $r$, we solve for state populations in the zero, one and multi-excitation manifolds. As decoherence only reduces populations in finite excitation manifolds, these results become the upper bounds for $P_1$ and $P_{\rm m}$ given by
\begin{eqnarray}
		P_1 & < & \frac{2e^{-rT}}{r^2-16\Omega_0^2}\Big\{ r^2 \cosh{\Big(\frac{T}{2}\sqrt{r^2-16\Omega_0^2}\Big)}e^{rT/2}  \nonumber\\
		& &  -16\Omega_0^2 e^{rT/2} - (r^2-16\Omega_0^2)\Big\}, \label{eq:P1}\\
		P_{\rm m} & < & 1-P_1 - {\rm exp}(-rT).	\label{eq:Pm}
\end{eqnarray}

In general, the optimal regime is accessible by using ultrashort and intense pulses, but such optical pumping may effect changes in the centers. In a femtosecond illumination experiment by Dumeige \textit{et al.}, they showed photochromism of a NV under a peak intensity of near $50$~GW/cm$^2$ with 55~pJ, 150~fs pulses~\cite{dumeige04}, and no damage to the diamond was reported. As similar studies of the NE8 and SiV are not present, we assume that these centers remain stable for a peak intensity $I < 50$~GW/cm$^2$.

Following Eqs.~\ref{eq:P1} and \ref{eq:Pm}, we choose parameters ($T,r$) = (0.16~ps, 20~THz) and (0.08~ps, 40~THz) for the NE8 and SiV respectively. These values correspond to $I =\hbar^2 c\epsilon_0 r^2/(2d^2) \sim 40$~GW/cm$^2$ and the energy per pulse is $\leq 50$~pJ. Consequently, emissions from the NE8 and SiV follow $P_1 < 0.97$ and 0.95 respectively, with $P_{\rm m} = O(10^{-5})$. 
On the other hand, for applications where minimizing $P_{\rm m}$ is favored over near unit $P_1$ (e.g., QKD in Sec.~III), shorter pulses can be used. For instance, for both centers, $20 - 40$~fs pulses are suitable for achieving $P_{\rm m} = O(10^{-7})$ for $P_1 \sim 0.56$. Development of modern mode-locking techniques made possible the generation of such ultrashort pulses (see Refs.~\cite{brabec00,baltruska02,keller03,cerullo03} for review articles on femtosecond lasers) with high repetition rates and notably, a titanium-sapphire laser capable of emitting 42~fs pulses at a 10~GHz repetition rate has been demonstrated~\cite{bartels08}. Gigahertz repetition rates can also be achieved by gating multiple ultrashort lasers.

\begin{figure}[tb!]
\includegraphics[width=1\columnwidth]{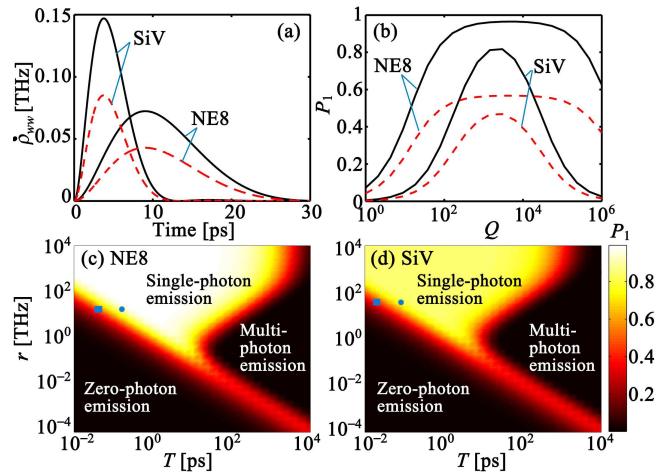}
\caption{(Color online) (a) Cavity-enhanced photon emission at ZPL wavelengths from two diamond optical centers: NE8 ($\lambda_c = 794$~nm) and SiV (738~nm). We take $V = (\lambda_c/n)^3$ and $Q$ = 3700 (NE8), 1800 (SiV). The solid curves are calculated using top-hat excitation parameters ($T,r$) = (0.16~ps, 20~THz) and (0.08~ps, 40~THz) for the respective centers. The dashed curves use shorter excitation pulses of $T$ = 40~fs (NE8) and 20~fs (SiV). Single-photon emission probability $P_1$ as a function of (b) $Q$, and (c,d) the excitation parameters. Circles (squares) indicate the parameters that effect $P_{\rm m} = O(10^{-5})$ [$O(10^{-7})$].}
\label{fig:pulses}
\end{figure}

Armed with this understanding, we present numerical simulations of a single excitation cycle. Figure~\ref{fig:pulses}(a) plots the temporal lapse and envelope of the outcoupled photon $\dot{\rho}_{ww} \equiv \langle g_0,0_c,1_w|\dot{\rho}|g_0,0_c,1_w\rangle$ in response to an excitation pulse that yields $P_{\rm m} = O(10^{-5})$. 
Excitations with shorter pulses that yield $P_{\rm m} = O(10^{-7})$ are also considered. 
The cavity-$Q$ values are modest: 3700 (NE8) and 1800 (SiV), and they are near optimal as shown in Fig.~\ref{fig:pulses}(b) where simulations are performed for different $Q$.
In the weak cavity regime of $\kappa > 2\Omega_0$, Purcell enhancement diminishes with $Q$. In the strong cavity regime where $Q$ exceeds $\omega_c/4\Omega_0$, the emission yield via the cavity channel reduces as the time scale for cavity relaxation becomes much longer than the other decay processes. 

The ratios $Q/V$ used in Fig.~\ref{fig:pulses}(a) imply emission enhancements with factors $F_p = 311$ (NE8) and 10 (SiV), and $P_1 = F_p/(1+F_p) = 0.997$ and 0.908 respectively if excitations are ideal. In the full treatment, we expect some reductions -- at large times the probability of emitting a photon of ZPL, given by $P_1 \equiv \langle g_0,0_c,1_w|\rho|g_0,0_c,1_w\rangle$, is 0.954 for the NE8. For the SiV, the effect of fast nonradiative decay $\tilde{\gamma}$ in the SiV is still significant but the cavity has improved $P_1$ from $0.05$ of a bare SiV to 0.812. When shorter excitation pulses are used, $P_1 = 0.565$ (NE8) and 0.470 (SiV). Irrespective of the pulse length $T$, the mean times for photon output are predicted to be 10~ps (NE8) and 5~ps (SiV), nearly three orders of magnitude faster than their nanosecond free-space lifetimes.

\begin{figure}[tb!]
\includegraphics[width=0.95\columnwidth]{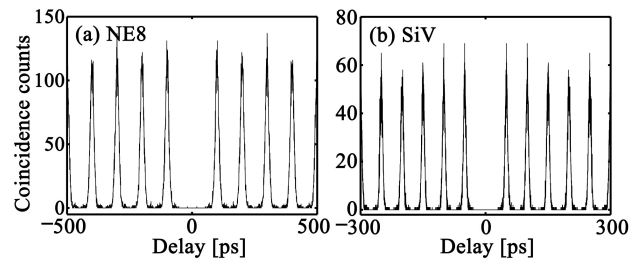}
\caption{Photon autocorrelation histogram of emissions from the cavity-center system implemented with an (a) NE8 and (b) a SiV center. Using quantum trajectory approach, HBT experiment is simulated using a pulsed excitation, corresponding to a top-hat pulse at respective repetition rates 10~GHz and 20~GHz for over 5000 emission cycles.}
\label{fig:HBT}
\end{figure}

\begin{table*}
\newbox\dothis
\setbox\dothis=\vbox to0pt{\vskip-1pt\hsize=11.5pc\centering
Error\vss}
\begin{tabular*}{\textwidth}{@{\extracolsep{\fill}}lcccccccccccc}\hline\hline
& & & \multicolumn{4}{c}{Cavity parameters} & \multicolumn{2}{c}{Excitation pulse} & \multicolumn{4}{c}{Source performance}\cr
\cline{4-7}\cline{8-9}\cline{10-13}
Optical  & Dipole & & Wavelength & Freq. & Optimal & Coupling & Pump & Pulse & Mean & Spectral & Single-ph. & Repet. \\
center  & mom.  & &  $\lambda_c$ & $\omega_c$ & $Q$ factor & $\Omega_0$ & rate $r$ & width & time & width $\Delta\lambda$ & prob. $P_1$ & rate $\nu$ \\
 & ($10^{-29}$Cm) & & (nm) & (PHz) & & (GHz) & (THz) & $T$ (ps) & (ps) & (nm) & & (GHz) \\\hline

\multirow{2}{*}{NV} &  $0.1^\ast$ & $\left(\bullet\right)$ & 637 & $2.95$ & 64000 & 9.3 & 1.3 & 2.3 & 125 & 0.007 & 0.95  & 2 \\
& $1^\dag$ & $\circ$ & -- & -- & -- & -- & 1.3 & 6 & $1.16\times10^4$ & $120$ & 0.74  & 0.086 \\\hline

\multirow{2}{*}{NE8} &  $2.1^\ast$ & $\left(\bullet\right)$ & 794 & 2.37 & 3700 & 127 & 20 & 0.16 & 10 & 0.15 & 0.954  & 30 \\
& $2.5^\dag$ & $\circ$ & -- & -- & -- & -- & $9$ & 1 & $1.15\times10^4$ & $2\times10^{-5}$ & 0.84  & 0.087 \\\hline

\multirow{2}{*}{SiV} &  $4.2^\ast$ & $\left(\bullet\right)$ & 738 & 2.55 & 1800 & 290 & 40 & 0.08 & 5 & 0.30 & 0.812  & 65 \\
& $4.7^\dag$ & $\circ$ & -- & -- & -- & -- & $10$ & 1 & $2.7\times10^3$ & $9\times10^{-5}$ & 0.05  & 0.37 \\\hline\hline
\end{tabular*}
\caption{Nominal parameters for single-photon generation with bare diamond optical centers [indicated by $\circ$] and cavity-enhanced centers [$\left(\bullet\right)$] with $V = (\lambda_c/n)^3$. Pulsed-excitation parameters are chosen so that multi-photon emission probability $P_{\rm m} = O(10^{-5})$. $\ast$ denotes the estimated dipole moment for the zero-phonon transition and $\dag$ the total dipole moment of the center across all phonon-assisted transitions.}
\label{tab:sum}
\end{table*}

The cavity-enhanced centers are now capable of operating at repetition rates of $\nu$ = 30~GHz (NE8) and 65~GHz (SiV). In these systems, multi-photon events are 4--6 orders of magnitude less likely than $P_1$, the probability of a dark interval per cycle is $O(10^{-4})$, and that of emitting a 1PL photon is $\langle g_1,0_c,1_w|\rho|g_1,0_c,1_w\rangle = O(10^{-6})$ (NE8) and $O(10^{-5})$ (SiV). Fourier transforms of the temporal profiles yield emission spectrums centered at the ZPL with effective linewidth of 0.15~nm (NE8) and 0.3~nm (SiV) at zero temperature. Although the pulse is non-Gaussian, techniques such as Stark tuning and $Q$-switching can be used to optimize pulse profiles~\cite{greentree06,fernee07,su08b,su09} and improve photon indistinguishability. 
Next the full model is solved for different excitation parameters in Figs.~\ref{fig:pulses}(c,d). The results are in agreement with Eq.~\ref{eq:P1} and show suitable ($T,r$) range for SPS. Although small improvements to the emission probabilities, $P_1 = 0.995$ (NE8) and $0.854$ (SiV) while maintaining $P_{\rm m} = O(10^{-7})$, are possible with $T\sim 10~$fs and $r\sim 1$~PHz pulses, the corresponding intensity is greater than those for observing photochromism.

Photon autocorrelation obtained in a Hanbury-Brown-Twiss (HBT) experiment is a test for single-photon emission. We simulate the experiment over 5000 excitation-emission cycles using the quantum trajectory approach~\cite{tian92,carmichael93}. Assuming the use of photon detection units with time bin sizes of 0.6~ps (for NE8) and 0.1~ps (SiV), the coincidence histograms $\propto g^{(2)}(\tau)$ in Fig.~\ref{fig:HBT} are compatible with perfect photon antibunching -- suppression of coincidence counts (i.e. multi-photon events) at and around zero delay.
We summarize the performance and implementation requirements of these centers, including the NV center from Ref.~\cite{su08}, in Table~\ref{tab:sum}. The assumed setup of ($T,r$)=(2.3~ps, 1.3~THz) implies a modest peak intensity 1~GW/cm$^2$ and pulse energy of 25~pJ on the NV, $P_1 = 0.95$, and $P_{\rm m} = O(10^{-5})$. Pulses of 6~pJ and $T = 0.6$~ps would give $P_1 = 0.54$ and $P_{\rm m} = O(10^{-7})$. We have also included the performance of the bare centers in the tabulation and their emission probabilities are estimated using $P_1 = 2\pi/(\tau\gamma_{\rm total})$.


\section{Quantum Key Distribution}
Robust and high-performance SPSs are an invaluable resource for implementing photon-based QKD protocols such as the BB84~\cite{bennett84}. In this section, we are interested in exploring the potential of diamond-based sources in a range of implementations by using BB84 as a metric for comparisons. Minimizing multi-photon emission probability $P_{\rm m}$ is fundamentally important for extending the range of secure communication, we therefore assume that the cavity-enhanced diamond centers are excited with suitable excitation $(T,r)$ (see Sec.~II) whereby $P_{\rm m} = O(10^{-7})$ at the cost of reduced $P_1 \sim 0.6$, and without further attenuation, $g^{(2)} = O(10^{-7})$. For the bare centers, we let $g^{(2)} = 0$. 

QKD allows two parties, Alice (sender) and Bob (receiver) to establish a secret key via a quantum channel that is open to an eavesdropper. Alice uses single-photon states to communicate classical information by encoding each photon in one of four polarizations: $0^{\circ}$, $45^{\circ}$, $90^{\circ}$, $135^{\circ}$ orientations. The first two correspond to classical bit 0 while the latter two bit 1. Alice (Bob) randomly chooses one of two bases, rectilinear or diagonal, to prepare (measure) the photons. Consequently, they communicate openly the bases they have used and retain those bits with matching bases. Finally, both parties perform local classical post-processing, namely error correction and privacy amplification, to correct erroneous bits and reduce the bit-string to a shared secure key asymptotically unknown to the eavesdropper. 

In the fiber-based scheme, time-bin encoding of signal states is preferred because single-mode fiber is susceptible to polarization loss. In contrast, polarization encoding is suitable for free-space QKD owing to the non-birefringent nature of the atmosphere. In both cases, the secure key generation rate $\mathcal{G}$ (bits~s$^{-1}$) of the protocol using a sub-Poissonian source is treated by Waks \textit{et al.} in Ref.~\cite{waks02},
\begin{equation}
	\mathcal{G_{\rm SPS}} = \nu q P_{\rm d} \left[ \beta\tau(e) - f(e)h(e) \right]
	\label{eq:GSPS}
\end{equation}
where $\nu$ is the repetition rate for bitstream single photons and $e$ is the total bit error rate. $q$ is the fraction of the correctly measured signals (typically $1/2$ because half the time Alice and Bob disagree on the bases chosen). The compression function $\tau(e)$ of the privacy amplification procedure is
\begin{equation}
  \tau(e) =  -{\rm log}_2\left[1/2+2e/\beta - 2(e/\beta)^2\right].
  \label{eq:error}
\end{equation}
$h(e)$ is the Shannon entropy function of a bit. $P_{\rm d}$ is the probability that Bob registers a detection event for a signal pulse. Although the mean number $\bar{n}$ of photons emitted from Alice's source is $(P_1 + 2P_{\rm m})\approx P_1$, she may add extra attenuation $\xi$ to reduce the relative number of multi-photon states. As a result, the key rate can be optimized with respect to this adjustable parameter $\xi$ for a given communication range. Therefore, the detection probability at Bob is $P_{\rm d} = \xi\eta P_1 + N$, where $\eta$ is the total optical loss in the system, $N$ is the total noise from the channel, the environment and the detection units.

In Eq.~\ref{eq:error}, $\beta$ is the fraction of detection events originating from single-photon states. Expressing in terms of the attenuation-free parameter $g^{(2)}$, the multi-photon emission probability becomes $\bar{P}_{\rm m} = \xi^2 P_1^2 g^{(2)}/2$ and $\beta = (P_{\rm d}-\bar{P}_{\rm m})/P_{\rm d}$~\cite{waks02}. The function $f(e)$ characterizes the performance of the error correction algorithm for a bit error rate $e = (e_0\xi\eta P_1 + N/2)/P_{\rm d}$, where the baseline error rate $e_0$ (typically 2\%) accounts for erroneous detection and state preparation. Signals can be lost during channel and apparatus transmission or coupling into and out of the channel, and due to detection inefficiency: $\eta = \eta_o\eta_c\eta_l\eta_d$, where $\eta_o$ (typically 60\%) is the total optical transmittance at Bob's end, $\eta_c$ the total coupling efficiency, $\eta_l$ the channel transmittance, and $\eta_d$ the quantum efficiency of a detector. 

For a quasi-SPS, $\mathcal{G_{\rm SPS}}$ scales with loss as $\eta$. When the upper bound on the allowable error rate is taken to be $\beta/4$, the loss cutoff for a sub-Poissonian source is~\cite{waks02}
\begin{equation}
	\eta_{\min} = \frac{1}{1-4e_0}\Big(\frac{N}{P_1} + \frac{P_1 g^{(2)}}{2}\Big).
\end{equation}
In other words, key generation between the parties remains possible if $\eta > \eta_{\min}$ so that $N$ and $g^{(2)}$ should be minimized for long-range communication.

On the other hand, the corresponding key rate formula for a weak coherent source (WCS) is~\cite{gottesman04}
\begin{equation}
	\mathcal{G}_{\rm WCS} = \nu q P_{\rm d} \left[\beta\tau(e/\beta) - f(e)h(e) \right]
	\label{eq:GWCS}
\end{equation}
where for a coherent state with mean photon number $\bar{n}$, $P_{\rm d} = 1-\exp(-\eta \bar{n}) + N$ when $\eta$ is small and noise level is the same for all photon states. The error rate $e = \{e_0[1-\exp(\eta\bar{n})] + N/2\}/P_{\rm d}$ and the probability of emitting a multi-photon state from Alice is $\bar{P}_{\rm m} = 1-(1+\bar{n})\exp(-\bar{n})$. Since the secure key is still extracted from the single-photon states, the source must be appropriately attenuated prior to transmission and the optimal photon number that maximizes the key rate is $\bar{n}_{\rm opt} \approx \eta$, leading to the scaling $\mathcal{G}_{\rm WCS} = O(\eta^2)$. The estimate for the corresponding loss cutoff is
$\eta_{\min} = \sqrt{2 N}/(1-4e_0)$~\cite{brassard00}. 

In the decoy state QKD, Alice uses additional vacua and multi-photon states to estimate the channel loss and error rate for each $n$-photon signal accurately~\cite{wang05,lo05} so that extra attenuation of the source is not required. The optimal photon number $\bar{n}_{\rm opt} \approx 0.7$ for $e_0 = 2\%$ so that the key rate is now of $O(\eta)$~\cite{lo05}:
\begin{equation}
	\mathcal{G}_{\rm WCS, decoy} = \nu q P_{\rm d} \left\{\beta [1-h(e/\beta)] - f(e)h(e) \right\}.
	\label{eq:GWCSdecoy}
\end{equation}

The noise level and optical loss vary greatly with specific QKD realizations. Here we explore three scenarios, namely QKD over (A) a silica fiber-optic telecom channel, (B) a free-space terrestrial channel, and (C) satellite-ground uplink and downlink. In each case we compare the performance realized using cavity-enhanced and bare diamond centers, with WCS counterparts at wavelengths 650~nm and 1.55~$\mu$m. 

The decoy state QKD protocol through a fiber link has been demonstrated by several groups. Rosenberg \textit{et al.} achieved 107~km range at 1.3~$\mu$m wavelength using high-performance transition-edge sensor (TES) photodetectors~\cite{rosenberg07}. Using the parameters in Ref.~\cite{gys04}, exchange over 140~km at 1.55~$\mu$m is feasible~\cite{lo05} with dark count rate of 100~Hz and 3.5~ns gate time. Dixon~\textit{et~al.} achieved 1~Mbit~s$^{-1}$ (10.1~kbit~s$^{-1}$) secure key rate over a telecom-fiber distance of 20~km (100~km) using an InGaAs photon detector at the same wavelength~\cite{dixon08}. Single NV centers in diamond nanocrystals have been used to realize secure exchange over short distances of 30--50~m in free space~\cite{beveratos02,alleaume04}. Long-distance exchange (over 144~km at 850~nm) has only been achieved using attenuated pulses and decoy states~\cite{sm07}. However, as we will show, while the decoy state protocol with high-speed WCSs are more suitable for practical QKD implementations than bare diamond centers, cavity-enhanced centers can deliver comparable performance and in some cases, outperform WCS-based implementations. We note that the feasibility and engineering of ground-satellite QKD system have been investigated by several authors~\cite{hughes00, rarity02, miao05, bonato09}. The first experimental study of such a setup has also been reported in Ref.~\cite{villoresi08}. 

\subsection{Silica-fiber based QKD}
In the fiber-based QKD scheme, the noise is largely due to dark counts at Bob's detectors. In a typical four-detector implementation, $N = 4R\Delta t$ where $R$ is the dark count rate and $\Delta t$ is the size of measurement gate window, usually chosen to match photon pulse length $\nu^{-1}$. Low dark counts have been achieved in various detection implementations. Widely available silicon avalanche photodiodes have a dark count rate $R \sim$ 25~counts~s$^{-1}$ and $\sim65$\% efficiency over the visible regime~\cite{SAPDdata}. Detectors well-suited for near-IR wavelengths are superconducting nanowire single-photon detectors (SSPD)~\cite{rosfjord06} and TES~\cite{rosenberg06}. At 1.55~$\mu$m, a TES has been shown to deliver 50\% quantum efficiency for an erroneous background count rate of 0.053~Hz whereas an SSPD can deliver 57\% efficiency with higher dark counts $\sim 1000$~Hz.

For practical distances, the dominant loss is due to fiber attenuation, given by $\eta_l = 10^{-\alpha l/10}$ where $\alpha$ is characteristic attenuation coefficient and $l$ is fiber length. The choice of silica-based optics favors the near-IR wavelength with $\alpha=0.2$~dB~km$^{-1}$, compared to $2.5 - 6$~dB~km$^{-1}$ in the visible. The bare centers suffer from low coupling efficiency into fiber modes such that $\eta_c \sim 1$\% for single photons. The directionality and coherence of the photons emitted via cavity mode can improve this efficiency to over 50\%~\cite{lipson04}. For the WCS, our choice of $\bar{n} = \bar{n}_{\rm opt}$ implies that the combined attenuation at Alice's end has already included WCS-to-fiber coupling loss so that $\eta_c$ for WCS can be set to unity. 

\begin{table}[tb!]
\begin{tabular*}{1\columnwidth}{@{\extracolsep{\fill}}lcccccc}\hline\hline
Source & & $\Delta t$ & $\alpha$ & $\eta_c$ & $N$ & $\eta_{\min}$ \\
 & & (ns) & (dB/km) &  & (1/pulse) &  (dB) \\\hline
\multirow{2}{*}{NV} & $\left(\bullet\right)$ & 0.5 & \multirow{2}{*}{6} & 0.5 & $5.0\times10^{-8}$ & 68 \\
    & $\circ$ & $420$ &  & 0.01 & $4.2\times10^{-5}$ & 28 \\\hline
\multirow{2}{*}{NE8} & $\left(\bullet\right)$ & 0.03 &  \multirow{2}{*}{2.5}  & 0.5 & $3.3\times10^{-9}$ & 74 \\
    & $\circ$ & 16.4&   & 0.01 & $1.6\times10^{-6}$ & 55 \\\hline
\multirow{2}{*}{SiV} & $\left(\bullet\right)$ & 0.02 & \multirow{2}{*}{3.5} & 0.5 & $1.5\times10^{-9}$ & 75 \\
    & $\circ$ & 3.4&  & 0.01 & $3.4\times10^{-7}$ & 50 \\\hline
WCS$^\ast$ &    & \multirow{2}{*}{0.1} & 6 & 1 & $1.0\times10^{-8}$ & 66 \\
WCS$^\dag$ &    & & 0.2& 1 & $2.1\times10^{-11}$ & 92 \\\hline\hline
\end{tabular*}
\caption{Nominal parameters for BB84, based on silica-fiber that is optimized for $1.55~\mu$, with different single-photon sources. The bare centers are indicated by $\circ$ and cavity-enhanced centers $\left(\bullet\right)$. $\ast$ denotes 10~GHz WCS at 650~nm and $\dag$ at 1.55~$\mu$m. The values of loss cutoff $\eta_{\min}$ for WCSs correspond to decoy state implementations.}
\label{tab:fibersum}
\end{table}

\begin{figure}[tb!]
\includegraphics[width=0.75\columnwidth]{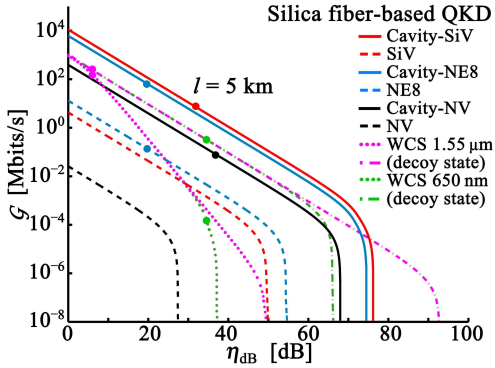}
\caption{(Color online) Secure key rate over silica-fiber link as a function of total loss, for different single-photon sources. Dashed: BB84 with bare optical centers NV (black), NE8 (blue), and SiV (red). Solid: BB84 with cavity-enhanced centers distinguished by same color curves. Dotted: BB84 with 10~GHz WCSs at 650~nm (green) and 1.55~$\mu$m (pink). Dashed-dotted: Decoy state with WCSs distinguished by same color curves. We use the parameters from Table~\ref{tab:fibersum}, $\eta_d$ = 0.65 (0.5), $R$ = 25~Hz (0.035~Hz) for visible (near-IR) wavelength transmission, $\eta_o = 0.6$ and $e_0 = 0.02$ for all cases. To aid comparisons, solid circles are used to indicate the key rates over fiber length $l$ of 5~km.}
\label{fig:qkdfiber}
\end{figure}

We summarize the key parameters and performance for the fiber-based systems with different sources in Table~\ref{tab:fibersum}. Equations~\ref{eq:GSPS}--\ref{eq:GWCSdecoy} are used to calculate the key rates as a function of total optical loss $\eta_{\rm dB} = -10\log_{10}{\eta}$ (i.e. in decibel) in Fig.~\ref{fig:qkdfiber}. The curves corresponding to WCSs are plotted with their respective optimal photon numbers. Here we compare transmissions clocked at the maximum repetition rate $\nu$ of the diamond-based SPSs with WCSs driven at rate 10~GHz. In practice, the repetition rate is limited by the timing jitter ($< 0.1$~ns) and reset time ($< 1~\mu$s) of the detectors but this effect can be ignored when these time scales are shorter than the mean arrival times of photons after a lossy channel. The prescriptions used here are still valid if any signals arrived during the reset time are discarded. To avoid fiber dispersion, only the emissions via ZPL transition of the bare centers are considered for QKD, and the ZPL spontaneous decay rates of the bare centers are estimated to be $2.5\times 2\pi$~MHz (NV), $60\times 2\pi$~MHz (NE8) and $300\times 2\pi$~MHz (SiV).

We obtain substantial improvements in terms of key rate (by 3--4 orders of magnitude) and range by using cavities. Over a 5~km fiber, the NE8 center delivers 62~Mbits~s$^{-1}$ key rate, the SiV 7~Mbits~s$^{-1}$ and the NV 0.08~Mbits~s$^{-1}$. However, since the telecom fiber favors near-IR transmission, a 1.55~$\mu$m WCS with decoy states is expected to be superior with over 200~Mbits~s$^{-1}$ key rate for $l = 5$~km. Given an ultralow noise level of $2.1\times10^{-11}$~pulse$^{-1}$ (i.e. 5 orders of magnitude less than that used in Refs.~\cite{gys04,lo05}), we estimate an optimistic 420~km communication range using the decoy states.

\begin{figure}[tb!]
\includegraphics[width=1\columnwidth]{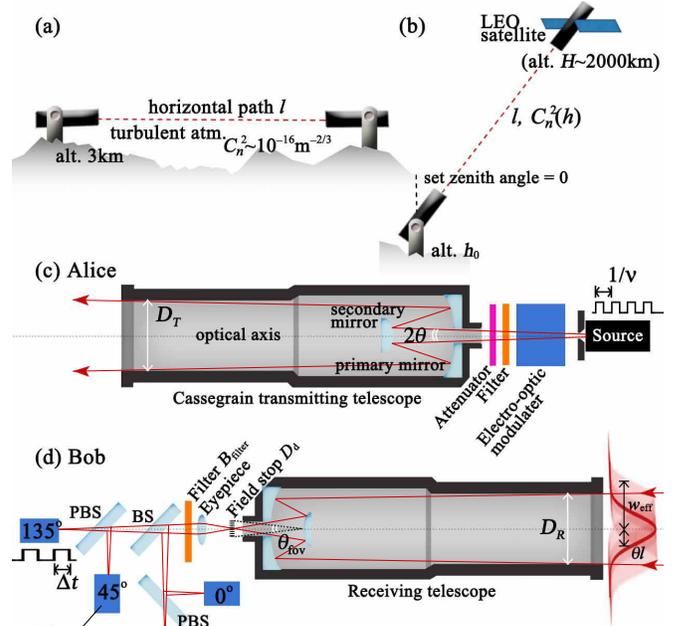}
\caption{(Color online) Generic (a) open-air terrestrial and (b) LEO satellite-ground communications over a path length $l$ through atmospheric turbulence characterized by refractive-index structure parameter $C_n^2$. (c) Transmitter using a typical Cassegrain telescope. Single photons generated at a repetition rate $\nu$ are encoded using electro-optic modulator. (d) Receiving telescope with an angular field of view $\theta_{\rm fov}$ sees a broadened spot with radius $w_{\rm eff}$. Measurement basis is randomly chosen with a beamsplitter (BS) and polarization states are distinguished with polarizing beamsplitters (PBSs). ATP system, post-processing, key management are not depicted here.}
\label{fig:openairschematics}
\end{figure}

\subsection{Open-air terrestrial QKD}
An open-air system is susceptible to stray photons and, in the visible and near-IR regimes, sunlight is the dominant source of such stray photons during the day, and starlight or artificial light sources dominate at night. The background brightness $H_{\rm b}$ observed from the ground varies from $1.5~{\rm W}{\rm m^{-2}Sr^{-1}\mu m^{-1}}$ on a clear daytime to $1.5\times10^{-3}{\rm Wm^{-2}Sr^{-1}\mu m^{-1}}$ on a full moon clear night~\cite{miao05}. However, the use of time-gate, frequency and spatial filters, together with an active high-precision acquisition, tracking and pointing (ATP) system between the transmitting (Alice) and receiving (Bob) telescopes, should improve the signal-to-noise ratio and allow communication over long distances~\cite{miao05}. Solar background can also be reduced by operating in the Fraunhofer windows. A setup is schematically depicted in Fig.~\ref{fig:openairschematics}. The gated noise counts per pulse is then given by~\cite{miao05}
\begin{equation}
	N = (H_{\rm b}\Omega_{\rm fov}A_{\rm rec}B_{\rm filter} + 4R)\Delta t, 
\end{equation}
where Bob uses a receiving telescope with a field of view (FOV) $\Omega_{\rm fov}$, a numerical aperture of diameter $D_R$ and area $A_{\rm rec}=\pi D_R^2/4$, gate sizes (temporal filter) of $\Delta t$ and frequency filters of width $B_{\rm filter}$. For a detector of diameter $D_d$ and a telescope with a focal length $f$ to aperture size ratio of $a = f/D_R$, the angular FOV $\theta_{\rm fov} = D_d/(D_R a)$ and the solid angle $\Omega_{\rm fov} = 2\pi(1-\cos{\theta_{\rm fov}}) \approx \pi\theta_{\rm fov}^2$ for a large $f \gg D_d$. When the telescope is used with an eyepiece, $\theta_{\rm fov}$ is the ratio of the diameter of the field stop to the focal length, as illustrated in Fig.~\ref{fig:openairschematics}(d).

Signal attenuation through the atmosphere is attributed to scattering by aerosols and absorption by atmospheric gases. 
In the visible and near-IR, the atmosphere is generally transmissive and the major gaseous absorber is water vapor. In particular, there are low-loss windows at around 770 and 870~nm~\cite{gisin02}. The typical values for attenuation by scattering at sea level can be found in Ref.~\cite{valley65} for a range of weather conditions and visibility, e.g., $\alpha \sim0.4-0.6$~dB~km$^{-1}$ ($\sim2.2$~dB~km$^{-1}$) for a visibility range of 23.5~km (5~km) in a standard-clear (medium-haze) condition. These values decrease with altitude, e.g., by one order of magnitude at $\sim3$~km above sea level. To estimate spectral transmittance of H$_2$O, one can refer to tabulations of experimentally determined transmission data (e.g. Ref.~\cite{passman56}) or the atmospheric model in the MODTRAN code~\cite{berk89}. As our reference we take the values corresponding to temperature 25$^\circ$C and relative humidity 60\%, the absorption loss is roughly $-12$~dB in the visible and $-4$~dB at 1.55~$\mu$m over a 150~km path at 3~km altitude.

The signal beam leaving the transmitting telescope diffracts and diverges to give rise to pulse spreading. For a Gaussian beam of wavelength $\lambda$, the divergence half-angle is $\theta = \lambda/(\pi w_0)$, where the radius of the beam waist $w_0$ should be narrower than half of the diameter $D_T$
of the collimating telescope (we let $w_0 = 0.35D_T$) so that most of the beam is transmitted. When the signal photons are detected at a distance $l$ away, Bob sees a broadened spot with diffraction-limited radius ($\sim \theta l$). Furthermore, atmospheric turbulence induces random displacement of the beam centroid (beam wander) when the beam propagates through optical turbules that are smaller than the beam diameter. This effect leads to a long-term beam radius~\cite{fante75}
\begin{equation}
	w_{\rm eff} = \theta l\sqrt{1+ (w_0/p)^2}, 
\end{equation}
where $k = 2\pi/\lambda$ is the wave number and $p = [0.55  k^2 l C_n^2]^{-3/5}$ for a flat profile of the refractive index structure along the propagation path. The typical values of refractive-index structure parameter $C_n^2$, which measures the strength of fluctuation in $n$, range from $10^{-17}-10^{-13}$ m$^{-2/3}$ under weak to stronger turbulent conditions. Thus the fraction of the light collected at the receiving telescope is
\begin{equation}
	\eta_c = 1- e^{-D_R^2/(2w_{\rm eff}^2)}.
	\label{eq:colleff}
\end{equation}

\begin{figure*}[tb!]
\includegraphics[width=\textwidth]{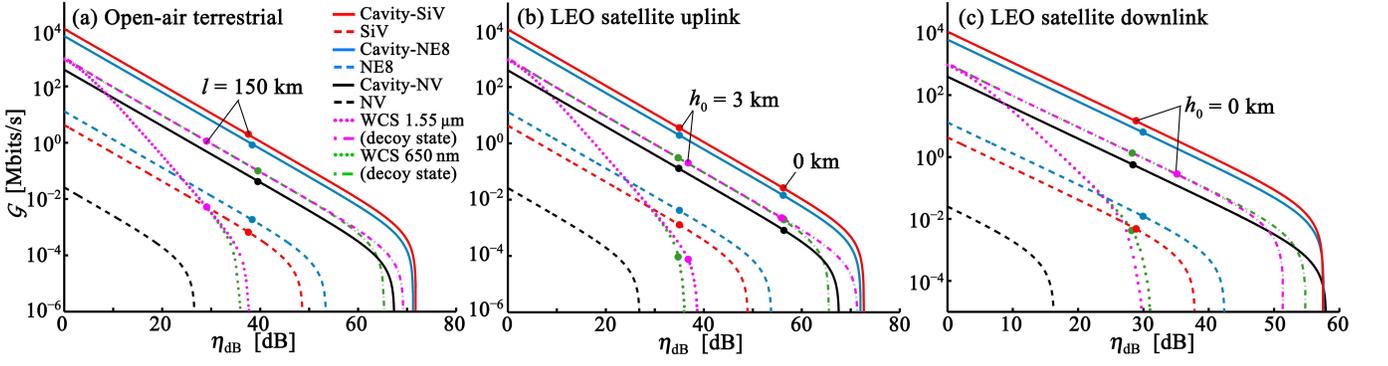}
\caption{(Color online) QKD performance of (a) open-air, terrestrial link at 3~km above sea-level, (b) free space uplink between a transmitter at an altitude $h_0$ and a LEO satellite at an altitude 2000~km, and (c) LEO satellite-ground downlink, on a standard-clear night with brightness $H_{\rm b}=1.5\times10^{-3}$Wm$^{-2}$Sr$^{-1}\mu$m$^{-1}$. In (a), $D_T = 0.15$~m at Alice, $D_R = 1$~m, $D_d = 0.5$~mm, $a = 39$, and FOV = 12.5~$\mu$rad at Bob. In (b), $D_T = 1$~m on the ground (Alice), $D_R = 0.1$~m, $D_d = 0.5$~mm, $a = 50$ and FOV = 100~$\mu$rad on the satellite (Bob). In (c), $D_T = 0.1$~m on the satellite, $D_R = 1$~m, $D_d = 0.5$~mm, $a = 5$ and FOV = 100~$\mu$rad on the ground. The values for $\eta_d$, $R$, $\eta_o$ and $e_0$ follow Fig.~\ref{fig:qkdfiber}, noise rates $N$ are tabulated in Table~\ref{tab:freesum}. Circles are used to indicate the estimated key rates for a particular transmission distance or altitude.}
\label{fig:fsoqkd}
\end{figure*}

\begin{table}[tb!]
\begin{tabular*}{1\columnwidth}{@{\extracolsep{\fill}}lc|cc|cc|cc}\hline\hline
&  & \multicolumn{2}{c}{(a)} & \multicolumn{2}{c}{(b)} & \multicolumn{2}{c}{(c)} \cr\cline{3-4}\cline{5-6}\cline{7-8}
Source &  & $N$ & $\eta_{\min}$ & $N$ & $\eta_{\min}$ & $N$ & $\eta_{\min}$ \\
 & & (1/pulse) & (dB) & (1/pulse) & (dB) & (1/pulse) & (dB) \\\hline
 \multirow{2}{*}{NV} & $\left(\bullet\right)$  & $6.0\times10^{-8}$ & 68 & $5.6\times10^{-8}$ & 68 & $6.1\times10^{-7}$ & 58 \\
  & $\circ$ & $5.0\times10^{-5}$ & 27 & $4.7\times10^{-5}$ & 28 & $5.1\times10^{-4}$ & 17 \\\hline
   \multirow{2}{*}{NE8} & $\left(\bullet\right)$  & $1.6\times10^{-8}$ & 72 & $1.1\times10^{-8}$ & 73 &  $7.4\times10^{-7}$ & 58 \\
  & $\circ$ & $2.0\times10^{-6}$ & 54 & $1.9\times10^{-6}$ & 55 & $2.5\times10^{-5}$ & 43 \\\hline
   \multirow{2}{*}{SiV} & $\left(\bullet\right)$ & $1.2\times10^{-8}$ & 73 & $7.9\times10^{-9}$ & 74 &  $6.3\times10^{-7}$ & 58 \\
  & $\circ$ & $4.2\times10^{-7}$ & 49 & $3.9\times10^{-7}$ & 50 & $4.8\times10^{-6}$ & 38  \\\hline
  WCS$^\ast$ & & $1.2\times10^{-8}$ & 65 & $1.1\times10^{-8}$ & 66 & $1.2\times10^{-7}$ & 54  \\
  WCS$^\dag$ & & $4.8\times10^{-9}$ & 69 & $2.9\times10^{-9}$ & 71 & $2.9\times10^{-7}$ & 51 \\ \hline\hline
\end{tabular*}
\caption{Estimated noise $N$ and loss cutoff $\eta_{\min}$ for (a) open-air terrestrial at 3~km altitude, (b) LEO (low earth orbit) satellite-ground uplink and (c) downlink implementation of the BB84 QKD protocol with different photon sources. We assume a night-time operation where the background brightness $H_{\rm b}=1.5\times10^{-3}$Wm$^{-2}$Sr$^{-1}\mu$m$^{-1}$. The values of $\eta_{\min}$ for WCSs correspond to decoy state implementations.}
\label{tab:freesum}
\end{table}

In the open-air terrestrial QKD setup, we consider a link at $\sim 3$~km above sea-level on a standard-clear, full moon night ($H_{\rm b} = 1.5\times10^{-3}~{\rm W}{\rm m^{-2}Sr^{-1}\mu m^{-1}}$, visibility $\sim$24~km). Following the Canary islands experiment reported in Ref.~\cite{sm07}, the emitted photons are collimated by 0.15~m optics and then intercepted by a 1~m diameter, $f/39$ telescope (39~m focal length) that has a FOV of 12.5~$\mu$rad for a typical detector size $0.5$~mm. We estimate a moderate $C_n^2 \sim 4\times10^{-16}$~m$^{-2/3}$ that causes a roughly $-14$~dB collection loss (i.e. $w_{\rm eff} \sim 3.5$~m) in that demonstration. Under these conditions, the magnitude of the beam wander is of same order as the diffraction-limited spot size at about 10~km (40~km) for the visible (1.55~$\mu$m) wavelengths, and beyond which beam-wander effect to dominate. Thus, while shorter wavelengths suffer less diffraction, this advantage diminishes for long distances $> 60$~km unless $C_n$ improves. For a typical picosecond time-of-arrival jitter, the setup can be clocked at the proposed repetition rates.
Given the availability of high-transmittance ($>90$\%) atom filters of width 0.01~nm, we set $B_{\rm filter} = 0.01$~nm or $\Delta \lambda$ of sources with wider spectral widths.

The estimated loss cutoffs of this setup with different sources are tabulated in Table~\ref{tab:freesum}, and the key rates are plotted in Fig.~\ref{fig:fsoqkd}(a). Over a 150~km link, NE8 and SiV centers can deliver a remarkable 0.8 and 1.8~Mbits~s$^{-1}$ respectively, whereas 10~GHz 650~nm and 1.55~$\mu$m WCSs with decoy states enables 0.1 and 1.2~Mbits~s$^{-1}$. The total losses of around $-35$~dB include an average $\eta_c = -14.5$~dB due to beam spreading, $-7$~dB due to atmospheric loss via scattering, $-4$ (near-IR) to $-12$~dB (visible) from H$_2$O absorption, and $-4.5$~dB due to apparatus imperfections. 
Since the loss cutoff $\eta_{\min}$ is around $-70$~dB during the night time, we predict over 300~km exchange is possible. On a clear day ($H_{\rm b} = 1.5~{\rm W}{\rm m^{-2}Sr^{-1}\mu m^{-1}}$), $\eta_{\min} \sim -46$~dB would still allow exchanges over distances $\leq 150$~km with the diamond-based sources.

\subsection{Satellite-ground QKD}
Under a standard clear-sky condition, transmission through the entire atmosphere between a ground-based station at an altitude $h_0$ and a satellite at $H = h_0 + l$ can be as high as 60--70\% at the visible and 85\% for 1.55~$\mu$m. Signals are largely lost at collection due to pulse spreading and beam wander. Along a zero-zenith angle path, the effective spot size is~\cite{andrews95} 
\begin{equation}
	w_{\rm eff} = \theta l \sqrt{1+7.75 \mu (k/\theta^5)^{1/3}},
	\label{eq:weffsatellite}
\end{equation}
where $\mu$ is related to varying strength of optical turbulence $C_n^2(h)$ as a function of altitude $h$. For uplink paths, 
\begin{equation}
	\mu = \int_{h_0}^H C_n^2(h)\Big(1 - \frac{h-h_0}{H-h_0}\Big)^{5/3} {\rm d}h
	\label{eq:muuplink}
\end{equation}
while for downlink paths,
\begin{equation}
	\mu = \int_{h_0}^H C_n^2(h)\Big(\frac{h-h_0}{H-h_0}\Big)^{5/3} {\rm d}h,
	\label{eq:mudownlink}
\end{equation}
and according to the Hufnagel-Valley atmospheric model,
\begin{eqnarray}
	C_n^2(h) & = & 0.00594(v/27)^2(10^{-5}h)^{10}{\rm e}^{-h/1000} + \nonumber \\
	         & & 2.7\times10^{-16} {\rm e}^{-h/1500} + C_n^2(0) {\rm e}^{-h/100}
\end{eqnarray}
where $h$ is in meters, $v$ is the pseudowind speed that ranges from 10 -- 30~m~s$^{-1}$ for moderate to strong wind, and nominal value of ground turbulence is $C_n^2(0) = 1.7\times10^{-14}$~m$^{-2/3}$. For satellite uplinks, beam wander is significant as the beam diameter is smaller than many of the turbules encountered. On the other hand, downlinks are less affected as the signal beam is already broadened by diffraction before entering the Earth's atmosphere.

Considering a satellite's allowed payload, we assume the use of 10~cm optics on the satellite~\cite{rarity02}. In the uplink communication between a low earth orbit (LEO) satellite at $H = 2000$~km and a ground-based telescope $h_0 = 0$ [Fig.~\ref{fig:openairschematics}(b)], Alice uses a 1~m diameter telescope ($w_0 = 0.35$~m) to transmit signal photons to $f/50$ downward pointing optics on the satellite with a 0.5~mm detector. This setup implies an effective beam spot radius of roughly 31~m for both visible and near-IR wavelengths, and collection loss (Eq.~\ref{eq:colleff}) of $-53$~dB. In general, $w_{\rm eff}$ decreases rapidly with increasing transmitter beam radius for small beams ($w_0 < 4$~cm) but this reduction becomes less pronounced for larger apertures ($w_0 > 4$~cm). The collection loss can be reduced significantly by situating Alice at higher altitudes, e.g., $h_0 = 3$~km yields $\sim$3~m beam at $H=2000$~km and $\sim -30$~dB loss.
Since the receiving optics has a wide FOV of $\theta_{\rm fov} = 100~\mu$rad, requirements on the tracking accuracy between the satellite and the ground station are not stringent.

Figure~\ref{fig:fsoqkd}(b) shows the key rates for operation on a standard clear night. While the setup is not optimized, we see that the centers can deliver up to 24~kbits~s$^{-1}$ key rate from the ground and over 3~Mbits~s$^{-1}$ from 3~km altitude, which are one magnitude faster than 10~GHz WCSs. The cavity-enhanced sources and higher altitude implementations also allow operations even in hazy daytime condition with $15~{\rm W}{\rm m^{-2}Sr^{-1}\mu m^{-1}}$ brightness and $\eta_{\min} \sim -40$~dB.

For the downlink case, the signal beam leaves the satellite with a spot radius of 3.5~cm and is taken to be detected by a $f/5$ 1~m telescope, where $w_{\rm eff} = 12$ (for 637~nm) to 28~m (1.55~$\mu$m) spot radius, or $-24$ and $-32$~dB losses respectively. Since the turburlence is much weaker at high altitudes, we expect from Eqs.~\ref{eq:weffsatellite} and \ref{eq:mudownlink} that the spot size is essentially the same as the diffractive spot size $w_{\rm eff} \approx \theta l$ at arrival. Thus the collection loss only improves significantly by using larger transmitting and receiving apertures, as opposed to using higher altitude receivers. Notably, it is advantageous to perform downlink at visible wavelengths because the divergence angle is larger at 1.55~$\mu$m, and this leads to a lower collection loss by near 8~dB. Fig.~\ref{fig:fsoqkd}(c) shows that the centers can produce up to $0.6-13$~Mbits~s$^{-1}$ key rates. Similarly, downlink can also be performed on a clear day with $\eta_{\min} \sim -29$~dB for the enhanced centers. Finally we summarize the expected noise levels and loss cutoffs for satellite-ground QKD at night time in Table~\ref{tab:freesum}.

\section{Conclusion}
We have studied the effect of a cavity on diamond defects, namely an NE8 and SiV optical centers, in enhancing its spectral properties for high-performance single photon generation. The Purcell enhancement in the high-$Q$ limit [$Q=O(10^4)$, $V=O(\lambda^3)$] of the weak coupling regime enables rapid photon generation on picosecond time scale. By applying suitable excitation schemes, these systems constitute high performance and fidelity single-photon sources that are capable of operating at $> 10$~GHz repetition rates. 

Standard BB84 using true single-photon sources provides an important benchmark implementation for quantum key distribution. In this work, we have contrasted bare and cavity-enhanced centers (including the NV center) with the weak coherent states in various prospective long-distance fiber-based and open-air setups. Fiber-optical communications with these diamond centers are expected to be short-ranged due to high loss through IR-optimized silica fiber at visible wavelengths. However, in open-air arrangements, cavity-centers can deliver performance that is comparable or better than the decoy state implementation. Notably, during day and night-time operation, these centers can achieve over 1~Mbits~s$^{-1}$ secure key rate in low earth orbit satellite-ground exchanges, and terrestrial point-to-point communications over 300~km path at 3~km altitude.

\section*{Acknowledgments}
We thank W. J. Munro, C. T. Chantler, A. Roberts, Z. W. E. Evans, A. M. Stephens, D. A. Simpson and B. C. Gibson for helpful discussions, and F. M. Hossain for the figures of the NE8 and SiV structures in Figs.~\ref{fig:schematics}(c,d). We acknowledge the support of Quantum Communications Victoria, funded by the Victorian Science, Technology and Innovation (STI) initiative, the Australian Research Council (ARC), and the International Science Linkages program. A.D.G. and L.C.L.H. acknowledge the ARC for financial support (Projects No. DP0880466 and No. DP0770715, respectively).

\end{document}